\documentclass[conference]{IEEEtran}
\IEEEoverridecommandlockouts

\usepackage{amsmath,amssymb,amsfonts}
\usepackage{algorithmic}
\usepackage{graphicx}
\usepackage{textcomp}
\usepackage{xcolor}
\def\BibTeX{{\rm B\kern-.05em{\sc i\kern-.025em b}\kern-.08em
    T\kern-.1667em\lower.7ex\hbox{E}\kern-.125emX}}

\usepackage{listings}
\usepackage{graphicx}%
\usepackage{multirow}%
\usepackage{amsmath,amssymb,amsfonts}%
\usepackage{amsthm}%

\usepackage{xcolor}%
\usepackage{textcomp}%
\usepackage{booktabs}%

\usepackage{listings}%

\usepackage[table]{xcolor}

\usepackage{tikz}
\usetikzlibrary{arrows.meta, positioning, fit}
\usetikzlibrary{calc,positioning, patterns,fit,arrows.meta,backgrounds,decorations.pathreplacing}

\usepackage[style=ieee]{biblatex}
\title{\textbf{Lost in the Pages}: WebAssembly Code Recovery
through SEV-SNP’s Exposed Address Space
}

\author{\IEEEauthorblockN{Markus Berthilsson}
\IEEEauthorblockA{
Markus.Berthilsson@eit.lth.se \\
Lund University\\
Lund, Sweden}
\and
\IEEEauthorblockN{Christian Gehrmann}
\IEEEauthorblockA{
Christian.Gehrmann@eit.lth.se \\
Lund University\\
Lund, Sweden}
}

\begin{document}

\maketitle

\begin{abstract}
WebAssembly (Wasm) has risen as a widely used technology to distribute computing workloads on different platforms. The platform independence offered through Wasm makes it an attractive solution for many different applications that can run on disparate infrastructures. In addition, Trusted Execution Environments (TEEs) are offered in many computing infrastructures, which allows also running security sensitive Wasm workloads independent of the specific platforms offered. However, recent work has shown that Wasm binaries are more sensitive to code confidentiality attacks than native binaries. The previous result was obtained for Intel SGX only. In this paper, we take this one step further, introducing a new Wasm code-confidentiality attack that exploits exposed address-space information in TEEs. Our attack enables the extraction of crucial execution features which, when combined with additional side channels, allows us to with high reliability obtain more than 70\% of the code in most cases. This is a considerably larger amount than was previously obtained by single stepping Intel SGX where only upwards to 50\% of the code could be obtained. 
\end{abstract}


\section{Introduction}
There is a clear trend towards dynamic usage of computing resources and networks. For instance, the sixth generation (6G) of wireless communication networks is expected to bring about a significant leap in the scope and capabilities of networked services. The networks are expected to offer ultra-high data rates, ultra-low latency, ultra-reliability, and energy efficiency \cite{Pennanen2025}. If it shall be possible to fully utilize the new capabilities, there is a need to also ensure efficient utilization of the network resources and the distribution of services. This requires platform independent solutions for workload orchestration and deployment \cite{VELASQUEZ2022311}. Wasm has arisen as one promising technology to achieve this and is finding usage outside the original browser environment \cite{Spies2021}. However, running security sensitive workloads in distributed infrastructures comes with confidentiality and privacy challenges. Usage of TEEs in the computing infrastructures is one natural direction to address these problems. TEEs are isolated computation units promising integrity and confidentiality, which attempts to eliminate this need of trust to the cloud service providers. Over the past years, large vendors such as AMD, Intel and ARM have each introduced their own implementations of TEEs \cite{kaplan2016amd,kaplan2017protecting,sev2020strengthening,costan2016intel,cheng2024intel,pinto2019demystifying}. Early TEEs were limited to processes, but since the release of AMD SEV, these TEEs have adopted a Virtual Machine (VM) based approached. This allows the customer to develop their system independent of the underlying hardware, and opens up the possibility to deploy TEE enabled VMs in cloud infrastructures. However, TEEs are designed with a harsh threat model where the hypervisor is considered to be malicious. With the high privileged mode, hypervisors are able to execute strong attacks using new controlled side channels.\\
Confidentiality in security has traditionally focused on ensuring that data remains accessible only to authorized users. While research mainly focuses on protecting input and output data, intellectual property today represents a massive financial investment. Consequently, the algorithms and executable logic themselves effectively become sensitive data, creating a critical need for code confidentiality. \\
In the context of cloud computing TEEs, code confidentiality promises that the software executed by a cloud service customer should remain secret to the cloud provider and any other adversarial party. This protection differs from data privacy, where not the data itself being hidden, but the executed instructions and logic. Recent work on code confidentiality \cite{puddu2024lack, lazard2018teeshift} focuses on the executed instructions themselves, rather than the confidentiality of the data that is processed by the code such as variables or keys. In particular, Puddu \textit{et. al} showed in \cite{puddu2024lack}, that it is possible to extract a large portion of the Wasm code by single stepping Intel SGX, and in particular considerable larger piece can be obtained compared to when doing the same attack on  native executed binaries.\\
In this work, we switch the focus to the AMD SEV-SNP TEE and present an attack on a Wasm interpreter, breaking code confidentiality inside the TEE. Wasm promises extra security in platforms by executing its modules in an isolated sandbox environment. A Wasm module consists of opcodes which are executed in a stack based VM managed by a runtime. However, these opcodes have to be translated to machine code before execution which results in multiple native instructions executed for every opcode. As a result, they are more susceptible for side-channel attacks. \\
Our attack is performed on a Wasm interpreter by analyzing the access patterns to memory space. More specifically, we show that the address space layout is a strong attack vector to extract more features for a side-channel attack. Through the use of these features, we were able to recover upwards of 70\% of the executed Wasm opcodes when running on WAMR in a AMD SEV-SNP enabled machine. \\
The main contributions of this paper are summarized as follows:
\begin{itemize}
\item We introduce a new recall metric for correct code confidentiality matching in TEEs. 
\item We present a partly new attack methodology for attacking Wasm code confidentiality in TEEs. In particular, we introduce a pre-processing step, where features such as page address of the Wasm stack and optable are extracted, which allows robust true operation code matches.
\item We demonstrate that with our attack method, we can perform very efficient code confidentiality attacks on WAMR running on AMD SEV-SNP obtaining more than 70\% of the code on average. 
\end{itemize}
We release all our code and data in an open repository\footnote{https://github.com/eDuuck/Sev-Wasm-Study}.\\ \\
This work is structured in 9 sections. First, section \ref{sec:related_works} highlights related works, followed by section \ref{sec:background} which summarizes a short background relevant for the study. Section \ref{sec:threat model} describes the attack model with section \ref{sec:problem_formulation} introducing the problem formulation. Section \ref{sec:methodology} explains the attack methodology with section \ref{sec:evaluation} containing an evaluation of the attack. Section \ref{sec:discussion} includes a discussion with future works and mitigations, and section \ref{sec:conclusion} concludes the content of the work.

\section{Related Works} 
\label{sec:related_works}
\textbf{Attacks on AMD SEV.} Many attacks have been performed on AMD SEV since its release in 2016. Early versions lacked CPU register encryption which allowed attacks that could recover plaintext arbitrarily from memory \cite{hetzelt2017security}.
AMD introduced SEV-ES to address this by encrypting CPU register states. However, SEV-ES did not ensure memory integrity which enabled attacks from the hypervisor by manipulating the nested page tables of the guest \cite{du2017secure,li2019exploiting,morbitzer2018severed,morbitzer2019extracting,wilke2020sevurity,morbitzer2021severity}. These attacks remap the memory of the SEV-ES enabled machine, causing the victim machine to either expose confidential data or execute unwanted code.\\
SEV-SNP is the latest iteration of SEV. While memory integrity has been introduced, multiple papers have shown how the platform is still susceptible to side channel based attacks. Weaknesses in the encryption method chosen in SEV have allowed for dictionary attacks \cite{li2021cipherleaks,li2022systematic}. Gast \textit{et al.} \cite{gast2025counterseveillance} showed that SEV-SNP still has as many as 228 performance counters exposed to the hypervisor, allowing a malicious hypervisor to infer the execution inside the confidential VM. From this they were able to fully recover RSA-4096 keys. These results show that SEV-SNP reduces the attack surface for host-controlled side channels but not entirely. \\

\textbf{Attacks on exposed address-space of TEEs.} Even when memory contents are encrypted, the sequence of memory accesses can reveal sensitive information. Xu \textit{et al.}\cite{xu2015controlled} showed on SGX, that by recording the sequence of page faults, an attacker could reconstruct input-dependent control transfers and data accesses. The authors demonstrated the severity of this attack vector by extracting full text documents and images processed by the machine.
Lee \textit{et al.} \cite{lee2020off} performed an a completely undetectable off-chip attack on SGX. By connecting external hardware to the memory bus, they were able to monitor memory accesses by the enclave. From this they were able to recover text document contents read inside the enclave. Similar to these works, our work uses the exposed address-space of the TEE to extract more side-channel features for our attack. \\

\textbf{Single-Stepping frameworks.} Many side-channel attacks on TEEs depend on the possibility to pause and inspect the victim's state at a granular level. The standard approach for this is to configure timer-based interruptions to preempt the victim's execution after a single instruction or basic block. This technique was pioneered for Intel SGX in SGX-Step \cite{van2017sgx}. An adaptation of the single stepping for AMD SEV-SNP  was presented in SEV-Step \cite{wilke2023sev}, and recently for Intel TDX in TDXDown \cite{wilke2024tdxdown}. We demonstrate our Wasm code confidentiality attack using the SEV-Step framework. \\

\textbf{Code confidentiality in TEEs.} Werner \textit{et al.} studied the leakage of instructions in the SEV TEE \cite{werner2019severest}. The lack of CPU register confidentiality in early SEV, and the possibility to trace guest execution through system calls, allowed the authors to gain critical information from the running guest without the need to single step it, but merely abort the guest at critical points by building application fingerprints. With this approach, they succeed both in breaking TLS obtaining the plaintext and injecting own keys for disc encryption for instance.  With the introduction of AMD SEV-SNP these attacks have been prevented.  Similar to the work in \cite{werner2019severest}, we also use instruction fingerprinting to build a reference database. However, our attack utilizes single stepping of the guest VM.\\
As far as we know, the only other paper addressing Wasm code confidentiality is the work by Ivan Puddu \textit{et al.} The authors investigated the instruction leakage of a Wasm runtime compared to native execution in Intel SGX by single stepping the enclave \cite{puddu2024lack}. The authors showed that the reduced instruction set of Wasm combined with the side channel amplification introduced by interpreters resulted in significantly less code confidentiality compared to native execution. They were able to extract upwards to 50\% of the instructions executed using side channels such as: memory access, branch records and instruction latencies. In this paper, we make a similar investigation for AMD SEV-SNP but expand upon it by utilizing memory address space patterns which shows considerable improved results compare to those obtained in Puddu \textit{et al.} for Intel SGX. \\

\section{Background}
\label{sec:background}
In this section we provide a brief technology background to the research area of our paper.\\

\begin{figure}[htb]
    \centering
    \resizebox{0.85\linewidth}{!}{
    \input{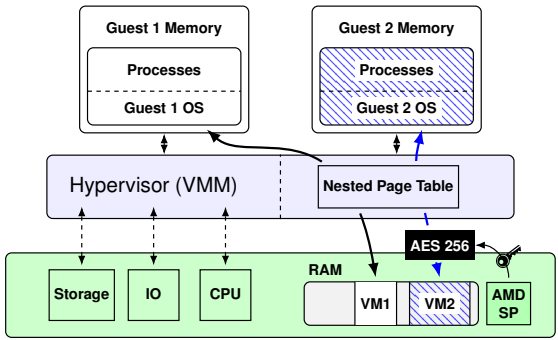}
    }
    \caption{Type-1 virtualization with two separate VMs. Guest 1 runs a machine without SEV-SNP enabled while guest 2 has it enabled. Dedicated hardware (AMD SP) manages keys for the individual VMs which adds confidentiality to the memory of guest 2. The hypervisor does not control the AMD SP and can not access the AES keys.}
    \label{fig:type1_virt}
\end{figure}

\textbf{Virtualization} is a fundamental technology for cloud computing which allows multiple isolated workloads to share the same physical hardware. A virtual machine (VM) behaves as an independent machine with its own operating system while executing on a shared host, with resource management performed by the hypervisor \cite{silberschatz2019}. The hypervisor manages memory translation of the VMs through Nested Page Tables (NPT) and dedicates hardware to the VMs. On AMD SEV-SNP, Type-1 (bare-metal) VMs can be run in a TEE, protected from the rest of the system as shown in figure \ref{fig:type1_virt}.\\

\textbf{Secure Memory Encryption (SME)} was introduced in 2016 to AMD's EPYC server processors. SME introduced a 128-bit AES engine onto the SOC which is able to perform encryption and decryption to the data that would reside in main memory. This defense was introduced to add confidentiality to the main memory in DRAM sticks which would defend against cold boot attacks \cite{halderman2009lest} or memory bus snooping based attacks. {AMD Secure Encrypted Virtualization (SEV) builds on top of SME, using the same AES engine but extends the security to multiple different VMs \cite{kaplan2016amd}. The CPU-integrated AMD Secure-Processor (SP) generates and manages the keys for the individual VMs. These keys are not accessible by the hypervisor.
SEV Encrypted State (SEV-ES) and \textbf{S}EV Secure Nested Paging (SEV-SNP) are two extensions to SEV that were released in 2017 and 2019, respectively. SEV-ES improved the confidentiality of the guest machines by also encrypting virtual CPU registers when the hypervisor raises a \texttt{VMEXIT} event, which halts the execution of the VM \cite{kaplan2017protecting}. 
SEV-SNP is the latest version of SEV which adds integrity of the guest memory \cite{sev2020strengthening}. AMD has proposed AMD Trusted I/O (AMD-TIO) to improve the security and performance with IO devices, released with the fifth generation of their AMD EPYC server CPUs.\\

\textbf{WebAssembly (Wasm)} is a stack-based binary instruction format introduced in major web browsers in 2017 \cite{haas2017bringing}. Its primary goal is to create a compilation target that is fast, secure and portable, making it well suited for deployment on the web. Wasm defines a set of low-level opcodes executed on a VM, ensuring that program execution is isolated from the rest of the machine. The 1.0 version of Wasm specifies 172 statically defined opcodes which operate on a stack. For example, an addition instruction (\textbf{i32.add}) pops the top two values of the stack and pushes the sum back on top of it. Wasm modules are either compiled from a programming language like C, or written directly with the Wasm semantics in a \texttt{wat} file.
Although Wasm was originally designed for web browsers, it's high performance, portability and security constraints has motivated the development of dedicated standalone runtimes which can be deployed outside of web browsers. Examples include wasmtime, wasmer, WAMR, WasmEdge and wasm3. In this study, WAMR was the picked runtime because of its low footprint and being highly modifiable. This made it into a good candidate for research and it has also been studied in other papers \cite{puddu2024lack}. Its core executable, \texttt{iwasm}, contains the interpreter used in our experiments. To keep track of the stack, \texttt{iwasm} keeps a stack pointer variable which points to to the top of the stack. After each opcode is executed, the interpreter fetches the next opcode and uses a lookup table to jump to the corresponding handler function. From here on, well refer to this lookup table as the \texttt{optable} as it plays a fundamental role in the attack methodology we propose.

\section{Attack Model}
\label{sec:threat model}
Our attack follows the AMD-SEV threat model \cite{sev2020strengthening} in which only the victim VM and the AMD hardware is considered to be trusted components. This is in line of assumptions made in previous related studies \cite{wilke2023sev}. The attack is carried out by a malicious or compromised hypervisor. As such, our attacker is able to make modifications to the Linux kernel and the attacker has full kernel-level privileges. This includes the ability halting and resuming the execution of a guest VM through the use of page faults or non-automatic exits. 

We assume a victim who deploys a confidential Wasm workload inside a SEV-SNP–protected VM, for example by running a standard open-source Wasm runtime. The attacker is assumed to be capable of identifying which runtime is in use, consistent with prior work showing that applications can often be fingerprinted inside TEEs via side-channel leakage  \cite{dipta2024dynamic,werner2019severest}. We assume that the runtime is used without code obfuscation or custom hardening. We work with the assumption of that a trusted entity is able to load the Wasm application into the TEE. For instance, through a secure channel. The provisioning can for example be done into a pre-configured and trusted cloud VM running on the TEE.

\section{Problem Formulation}
\label{sec:problem_formulation}
Under the previous given threat model, we are interested to investigate how well an external attacker with full control of the hypervisor is able to break the AMD SEV-SNP code confidentiality. We seek suitable side channels to utilize and with experimental evaluations investigate how large portions of Wasm code it is possible to extract. The instruction set of an interpreter like Wasm is much smaller than the x86-64 instruction set, consisting of only 172 core instructions. Every single instruction that has to be parsed and executed by a Wasm interpreter has to perform multiple x86-64 instructions, amplifying the side-channel leakage. Figure \ref{fig:wasm_i32_add} shows as an example how the WAMR interpreter amplifies the leakage of a simple (\textbf{i32.add}) instruction to 8 assembly instructions. This trace of instructions will cause memory access patterns which also leak information about the parsed opcodes.

\begin{figure}[htbp]
\centering
\begin{tabular}{|l|}
\hline
\textbf{C Code} \\
\hline
\begin{lstlisting}[basicstyle=\ttfamily\footnotesize,aboveskip=-5pt, belowskip=0pt,numbers=none]
HANDLE_OP(WASM_OP_I32_ADD)
{
    frame_sp -= 1;
    *(frame_sp-1) += *frame_sp;
    goto *handle_table[*frame_ip++];
}
\end{lstlisting}\\
\hline
\textbf{x86-64 Assembly} \\
\hline
\begin{lstlisting}[basicstyle=\ttfamily\footnotesize,aboveskip=-5pt, belowskip=0pt,numbers=none]
endbr64                          
sub     rbx,     4
mov     eax,     [rbx]
add     rbp,     1     
add     [rbx-4], eax
movzbl  eax,     [rbp-1]
mov     rax,     [r10+rax*8]
jmp     rax
\end{lstlisting}\\
\hline
\end{tabular}
\caption{WAMR Classic Interpreter handler for \textbf{i32.add} in C source code and resulting x86-64 assembly output after compilation.}
\label{fig:wasm_i32_add}
\end{figure}

\section{Methodology}
\label{sec:methodology}
Our goal is to determine the Wasm code for an unknown application. We approach the problem by performing an attack in two phases; First a profiling phase where known code is run by the attacker on an identical system which is going to be attacked later. The second phase is the attack phase where victim execution is intercepted, and the confidential code is comprised based on the result from the first phase. Each of these phases are carried out in 3 steps. The first two steps are identical in both the profiling and the attack phase while the third step differs. The first step is to collect side-channel data by single-stepping a VM through the use of the SEV-Step (see section \ref{sec:related_works}). In the second step the data is pre-processed and features are extracted from the side-channel data. The final third step differs for the profiling and attack phase. In the profiling phase, the third step consist of building a database of fingerprints of the different Wasm opcodes. In the attack phase, we match the extracted data to the database constructed in the profiling phase. The rest of this section describes these four different steps in more detail. 

\subsection{Data Collection}
\label{sec:methodology_data_coll}
We use the SEV-Step framework which consist of a custom Linux kernel with a modified Kernel Virtual Machine (KVM) module. The stepping of the machine is achieved by by utilizing Non-Automatic Exits (NAE) which triggers a \texttt{VMEXIT} in the victim VM, after which data can be collected for side channel analysis. The NAEs are timed with an Advanced Programmable Interrupt Controller (APIC) to occur just as the VM resumes its execution. This allows the VM to fetch and execute a single instruction before managing the \texttt{VMEXIT}. A spin-lock keeps the machine frozen until the hypervisor has performed the desired actions such as performing measurements or modifying the nested page tables. 

As single-stepping slows down the machine significantly, it should only be performed when the machine is executing the targeted interpreter. One way to achieve this is through page-faults which are used to track the access patterns of the victim to set appropriate breakpoints of the measurements. Prior works \cite{li2019exploiting,werner2019severest,dipta2024dynamic} has shown how side-channel information can be used to fingerprint applications to great accuracy.

Once the \texttt{iwasm} binary has been loaded, single-stepping is enabled. Side-channel information is collected until the interpreter finishes the Wasm module. After each step, instruction latency and memory interaction is recorded as side-channel data. Works such as \cite{van2018nemesis,puddu2021frontal,uops} has shown how different x86 instructions have different timing characteristics which leaks information about the system. This latency is measured as the time it takes from the NAE to be triggered until the machine has been paused. After each single-step, access rights to all the victim memory pages are removed. When execution of the victim machine is resumed, multiple page-faults will be raised as the CPU accesses different memory regions in preparation to execute the next instruction. These page-faults reveal if the fetched memory location will be executed (E), written to (W) or read (R). In our side-channel collection, we store the amount of page-faults an instruction raises, the type of page-fault, page address and instruction latency.
\subsection{Data Pre-processing}
\label{sec:methodology_preproc}

After collecting the side channel data through side-stepping, we pre-process the data to remove redundant data and extract important features in the measurement. As mentioned in section \ref{sec:background}, \texttt{iwasm} uses an \texttt{optable} to get the correct handler function for each Wasm instruction. The first step of pre-processing is to extract the page location of the \texttt{optable}. Once the \texttt{iwasm} interpreter fetches an instruction, it uses the \texttt{optable} to look up the corresponding handler function before executing the jump instruction. This creates a characteristic \textit{read-then-execute} pattern that appears regularly in in the side-channel trace. By filtering the collected data for this pattern, we can reliably extract the page containing the \texttt{optable}. In all of our experiments, we were able to extract the correct page corresponding to the \texttt{optable} using this approach.\\
The stack can be extracted in a similar manner. Now pages are filtered by picking pages that are often read and written to. Since the \texttt{optable} and stack pages are accessed in close proximity, we pick the best set of potential stack pages that ensure that most regions with \texttt{optable} accesses also contain at least one of the stack page candidates.\\
Finally we can remove redundant data such as operations done outside of the interpreter binary. This is done to speed up the matching process in the attack phase.

\subsection{Profiling}
\label{sec:methodology_profiling}
After collecting and pre-processing data from known execution in a SEV-SNP enabled machine, we match segments of the side-channel data to Wasm opcodes. The goal is then to build up a database of traces that will be used to fingerprint the Wasm opcodes which will later be matched against executions of unknown Wasm modules inside the trusted execution environment.
We began by constructing a reference Wasm module from a \texttt{wat} file containing all instructions defined in the Wasm MVP specification. This module was then loaded into a modified WAMR runtime, running inside a VM protected by AMD SEV-SNP. The modification introduced an additional memory write to a known pre-allocated page before the fetch of each new Wasm instruction. The addresses from the page faults reveal which events relate to these extra instructions, which then splits the captured trace up into segments. The segments were then matched to the known Wasm execution, creating a database of instruction fingerprints as illustrated in figure \ref{fig:profiling}. This method of profiling would be applicable to other runtimes and TEEs as well, as long as page-faults can be abused by the attacker. All that is needed is to introduce a dedicated page-write in the attacked runtime, prior to fetching the each new instruction.
\begin{figure}[ht]
    \centering
    \resizebox{\linewidth}{!}{
        \input{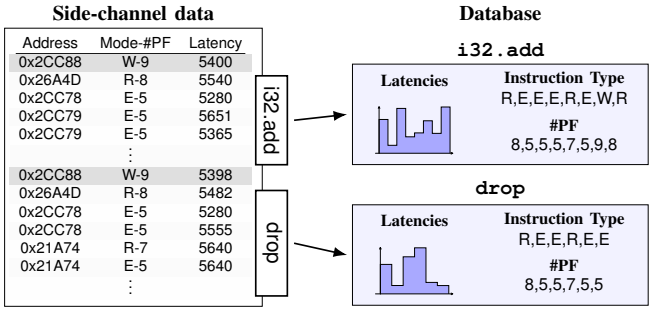}
    }
    \caption{Process of profiling the victim machine. After collecting a trace of side-channel data, segments are spliced from the side-channel data with the introduced writes, highlighted in gray. The following segments are then matched to individual Wasm instructions which are stored in a database.}
    \label{fig:profiling}
\end{figure}

To establish a ground truth, we executed the same module on an unmodified WAMR instance to verify our trace segments. Additional traces were collected from Wasm modules compiled from small C programs, broadening the instruction coverage and contextual diversity of our dataset.
After collection, the raw fingerprints were filtered to remove redundant entries. Specifically, fingerprints exhibiting identical page-access patterns, trace lengths, stack behaviors, and other structural traits were grouped together. Each group’s latency measurements were then averaged to produce a single representative fingerprint, yielding a clean and consistent dataset suitable for subsequent analysis.

\subsection{Matching}
\label{sec:methodology_matching}
In the attack phase, the executed opcodes of the victim machine is unknown. Traces are first captured and pre-processed in the same steps as in the profiling phase. As mentioned earlier, the following side channels are collected: instruction latency, instruction type, amount of page-faults per instruction and page addresses. From the page addresses we extract the stack pages and the \texttt{optable} page in pre-processing. The steps of the attack phase is illustrated in figure \ref{fig:matching_fig}.
\begin{figure*}[htpb]
    \centering
    \resizebox{\linewidth}{!}{
        \input{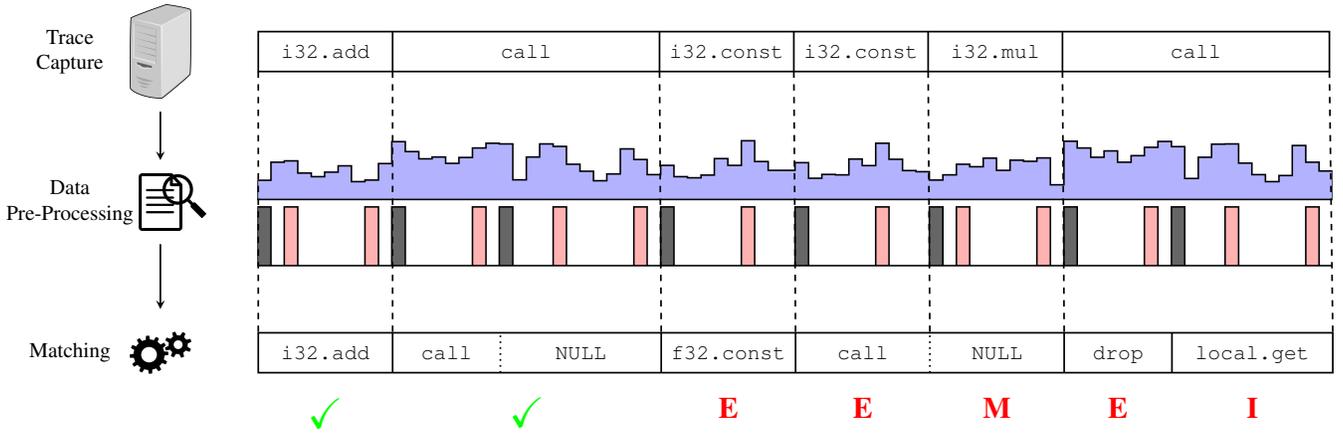}
    }
    \caption{The three stages of attack phase. 
            \textbf{1.} Data is captured when a SEV-SNP enabled machine is running a trace of Wasm instructions. 
            \textbf{2.} The data is preprocessed, top graph representing the latencies of individual x86 instructions and the bottom graph containing gray lines for when the \texttt{optable} page is accessed and red lines when operations modify the stack. 
            \textbf{3.} The matching algorithm extracts instructions based on the preprocessed side-channel data. Either getting a correct output (\checkmark), error (E), miss (M) or insertion (I).}
    \label{fig:matching_fig}
\end{figure*}
When the algorithm evaluates one or multiple segments, it calculates a score for all the fingerprints in the database. For each side-channel trace in a segment, a score between zero and one is calculated. The final score is then the sum of multiplying these scores together, with a maximum value of one and a minimum value of zero.\\
For numerical side channels, such as latencies and the amount of page-faults from a single instruction, Pearson correlation coefficient is used:
\begin{equation}
    r_{xy} = \frac{\sum^n_{i=1}(x_i-\overline{x})(y_i-\overline{y})}{\sqrt{\sum^n_i(x_i-\overline{x})^2}\sqrt{\sum^n_i(y_i-\overline{y})^2}}
\end{equation}
Where $n$ is the segment length, $x$ is the measured data from the attack and $y$ is the data in the fingerprint. $\overline{x}$ and $\overline{y}$ are the average over the sample size.\\
Discrete side channels use a use hamming distances to calculate a trace score. This includes: instruction type (R,W,E), stack access and \texttt{optable} page accesses. As the hamming distance is unbounded growing function, we used the following equation to limit the score between zero and one:
\begin{equation}
    s_s = \frac{1}{1 +\sum_{i=1}^n1(x_i\neq y_i)}
\end{equation}
Where $s_s$ is the score from a single discrete side-channel and $\sum_{i=1}^n1(x_i\neq y_i)$ is the hamming distance between the two input vectors $x$ and $y$. Once all the individual scores for each side channel is calculated. The final score is the multiplied sum of the individual side channel scores. The best matched fingerprint is picked and the corresponding Wasm opcode is extracted. \\
While there may be better ways to calculate the fingerprinting score, we found that this method were robust and performed well. Any function that inputs two vectors and outputs a scalar between zero and one could be used for the matching algorithm.

\section{Evaluation}
\label{sec:evaluation}
To evaluate the effectiveness of our attack we need to define a metric of the performance. In the matching phase, our script will select a region of our pre-processed data based upon \texttt{optable} accesses. From this region it will compare all the instructions from the fingerprint database and pick the best match. If the extracted opcode matches the executed opcode by the victim machine, we have an correct outcome. However, some instructions like \textbf{call} and \textbf{memory.grow} accesses the same page as the \texttt{optable} is stored in, resulting in multiple accesses to that page. Since our algorithm uses the accesses to the \texttt{optable} page to split our measurement file, this would result to multiple opcodes extracted from a single opcode. To remedy this, we insert a \textit{NULL} opcode in this region to indicate the lack of a new instruction as illustrated at the bottom row of figure \ref{fig:matching_fig}. \\
Thus, there are four different possible outcomes from the extraction of a region. Aside from the correct extraction, we can have an error where the extracted opcode does not match the executed opcode. If the algorithm extracts a \textbf{call} the following region will result as a \textit{NULL} opcode. If this \textbf{NULL} instruction region actually corresponds to a different instruction, then the algorithm has missed an instruction. Similarly, if the machine executes a \textbf{call} and the algorithm extracts another opcode, then an insertion will be made on the following region. Thus, we use a recall metric for our evaluation defined as:

\begin{equation}
    r_c = 1 - \frac{E + M + I}{N}
\end{equation}

Where \textit{N} is the total number of extraction regions. \textit{E} is the total number of wrongly evaluated instructions, \textit{M} are the missed instructions and \textit{I} is the inserted instructions.\\
To justify this metric, consider a side-channel attack extracting the following text from the secret below:
\begin{center}
\texttt{"Side\textbf{a}-Ch\textbf{o}nnels ar\textbf{a} interestin\char32"}\\
\texttt{"Side-Channels are interesting"\textcolor{white}{\char32}}\\
\end{center}
A naive approach to compare the output character one to one, would result in a low recall metric of 17\%, even though most of the data was perfectly reproduced. However, due to the noisy nature of side-channel attacks, an early insertion in the word \texttt{"Side"} caused the entire rest of the output to be shifted. Using our metric instead, two errors are present, one missed character, and one inserted character. This results in a much more fair recall metric of 87\%.\\
Finally, we group up instructions that are functionally identical with different byte sizes. \textbf{i32.add} and \textbf{i64.add} are considered the same instruction in our evaluation where only the byte-size of the opcode differs. Extracting either of these will result in a correct guess in our evaluations. This was done in \cite{puddu2024lack}, and as such for comparison sake, we do the same.

\subsection{Experimental Setup}
\label{sec:evaluation_exp_setup}
We evaluated our attack on a machine featuring an AMD EPYC 8124P processor, running the latest version of AMD SEV-SNP. The machine was running Ubuntu 22.04 with the victim guest VM running the same operating system. In the machine we ran the SEV-Step kernel with some minor modifications. We found that calling a Table Lookaside Buffer (TLB) flush was not possible once a VM was halted from single-stepping. When SEV-Step modifies the NPTs to introduce page-faults, a TLB flush is called at the end. This caused the machine to freeze when trying to modify the NPTs. As such, we added a tweak which removes the TLB flush if single-stepping was enabled.

\subsection{Results}
\label{sec:evaluation_results}
\begin{table*}[htbp]
\centering
\resizebox{0.7\linewidth}{!}{
\rowcolors{3}{gray!25}{white}
\begin{tabular}{c|r|ccc|c|c}
\toprule
\textbf{Module} & \textbf{Extracted Ins} & \textbf{Errors} & \textbf{Misses} & \textbf{Insertions} & \textbf{Recall} & \textbf{Script Runtime [s]} \\
\midrule
    AES O0     & 1,320,561   &   306,935 &   328  &   139 & 76.722\% &  678.55   \\
    AES O1     &    69,444   &    20,389 &     7  &     0 & 70.630\% &   34.86   \\
    AES O2     &    44,351   &    14,269 &    78  &     1 & 67.649\% &   22.08   \\
    primes.wat &   213,900   &    44,032 &  2,870 &     1 & 78.072\% &   98.97   \\
    Chess O0   & 8,568,798   & 2,767,703 & 21,143 & 3,988 & 67.407\% & 6136.69   \\
    Chess O2   & 1,905,184   &   543,264 &  1,018 &   690 & 71.395\% & 1278.79   \\
    Chess O3   & 126,516,521 &32,598,993 & 92,605 &52,091 & 74.119\% & 143297.74 \\
\midrule
\end{tabular}
}
\caption{Results of the attack on 7 different Wasm modules compiled from different targets.}
\label{tab:byte_ex}
\end{table*}
We evaluated our attack on seven Wasm modules performing three different workloads: a chess engine, AES-128 encryption/decryption from tinyAES, and a simple hand-written prime-number calculator from a  \texttt{wat} file. The chess engine was compiled with optimization levels \texttt{O0}, \texttt{O2}, and \texttt{O3} from an open source engine written in C. For \texttt{O0} and \texttt{O2} the search depth was restricted to one, while the \texttt{O3} build evaluated a higher search depth of three. The AES module was compiled with \texttt{O0}, \texttt{O1}, and \texttt{O2} levels. All modules were executed inside the \texttt{iwasm} interpreter under SEV-SNP, and instruction traces were reconstructed using the extraction and matching procedure described previously.

Table \ref{tab:byte_ex} summarizes the results. The consistency of our results across different workloads shows that our extraction method works well disregarding of code size, control-flow structure and compile optimizations. Across all binaries, the attack achieves a correctness between 67\% and 78\%, substantially outperforming prior work such as Puddu \textit{et al.}~\cite{puddu2024lack}, who reported 28\% recovery inside from the same \texttt{iwasm} interpreter under SGX. That attack also relied on last branch records, from which they could infer if the instruction would be a control flow instruction. Our main advantage is the use of the address space, from which we extract more features such as stack pages and the \texttt{optable} page. Their attack relies on using regex expressions to match the traces, while our matching method is more robust, using a scoring metric for the different fingerprints to extract the best fit for a side-channel data segment.

The hand-written \texttt{primes.wat} module yields the highest correctness (78.0\%). Its simple execution loop keeps the entropy of the executed Wasm module low resulting in predictable outputs. We looked into the execution trace, and only 17 unique opcodes were used for the execution. The AES modules on the contrary used 30-33 unique opcodes depending on the compiler flags, while the chess engine used 47-48 unique opcodes.\\
The chess engine presents a more complex case, especially in the \texttt{O3} configuration, where deeper search leads to substantially longer traces (over $10^8$ extracted regions). Despite this increased complexity—and the corresponding growth in error, miss, and insertion counts—the correctness remains comparable to the simpler workloads. This indicates that the attack remains robust even for long-running, branch-heavy workloads.

Overall, the results demonstrate that our approach is capable of recovering a substantial fraction of the executed instructions across diverse Wasm workloads and significantly improves upon prior instruction-recovery attacks in trusted execution settings.

\begin{figure}[htbp]
    \centering
    \includegraphics[width=\linewidth]{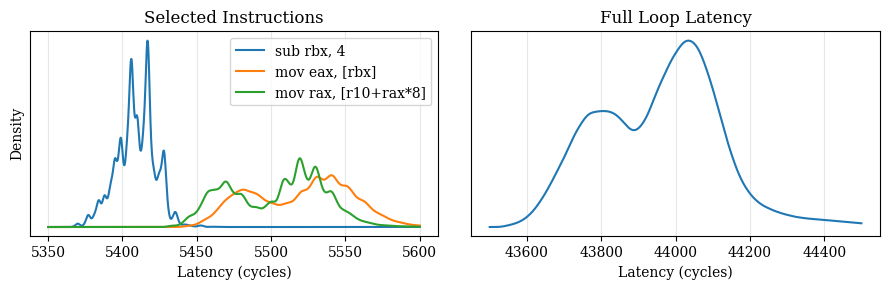}
    \caption{\textbf{Left}: Latency density of three different instructions, each one measured 1M times.
             \textbf{Right}: Collective latency density of 1M \textbf{i32.add} opcodes executed.}
    \label{fig:latencies}
\end{figure}

There are several factors that contribute to the errors in our measurements. Firstly, the latency measurements are noisy as both the \texttt{VMEXIT} and APIC timer used for single-stepping introduce sources of noise. To illustrate this, we collected a trace of a tight loop executing the same instructions as the opcode \textbf{i32.add} one million times. Figure \ref{fig:latencies} show how the latencies of single instructions differ. While instructions that perform read or writes to memory are substantially slower than instruction performed on registers, there is still overlap of the instructions. The latency curve for the \textbf{sub} instruction shows how the low resolution of the APIC timer offsets the latency measurements, causing spikes in the density curve. Across the 8 instructions performed for \textbf{i32.add} (see in figure \ref{fig:wasm_i32_add}), this introduces a large spread of collective latency between 43500-44600 cycles. This noise affects the score calculated by the Pearson Correlation Coefficient which introduces errors. \\
A second major contribution to noise is system interference. Even in our tightly controlled loop of 10 million assembly instructions collected for figure \ref{fig:latencies}, we observed 1953 context switches, adding 4,409,447 extra instructions to our loop. All of these context switches add uncertainty to our traces, lowering the accuracy of our script. Finally, due to the inaccuracy of the APIC timer, it is possible to miss instructions when attempting to single-step. However, in our largest trace from the \texttt{O3} compilation of the chess engine, we only observed 10 multi steps across 2,810,963,156 measurements. As such, the errors introduced due to multi-steps are negligible.

\section{Discussion and Future Works}
\label{sec:discussion}
Our findings show that the that the address space poses as a large vulnerability when running an interpreter inside AMD SEV-SNP. Even though our side-channel extraction was less feature rich than the similar work done on SGX \cite{puddu2024lack}, we were able to recover substantially more confidential code through a single trace capture of a victim machine. This is in large part due to the pre-processing step where more features are extracted such as the page addresses of the stack and optable. The pre-processing allows us to also use strong numerical methods to match our opcode fingerprints such as the Pearson correlation. This makes our matching method robust to noise in the latency measurement, unlike regex matching method used in the previous work on SGX. Actually, we found in our experiments that using correlation metrics alone on the latencies were enough to achieve recall values around 60\%. This would make it seem that the latency side channel is alone responsible for our results. But running the matching algorithm without using the Pearson correlation metrics, we found that the recall metric only reduced by 8-15\%. This indicates that memory-access patterns alone, without timing information or single-stepping, already leak enough structure to reconstruct a significant portion of the executed opcodes.\\
A key practical limitation of single-stepping in SEV is its extreme overhead. From some quick measurements, we found that single-stepping approximately seven million instructions slowed execution from 22.2 ms natively to 38.666 seconds, a 17 405x slowdown. This makes the attack very detectable if any external monitoring of the system is present. However, as address space revealed by page faults is sufficient to recover a substantial amount of the executed opcodes, undetectable off-chip attacks \cite{lee2020off} may be feasible. \\
While this work utilized address space to extract more features of the side-channel data, there's still much room for improvements. The memory layout the interpreter remains unused in our matching pipeline. Once the VM has loaded the interpreter into main memory, the layout remains the same throughout the lifetime of the process. Our collected fingerprints reveal that frequently used instructions such as \textbf{i32.add} and \textbf{f32.add} consistently reside on different pages across all tested configurations. Even a simple “execution-page” constraint could therefore significantly limit the space of possible opcodes for a segment of  measurement data. Similarly, we did not deploy cache-based techniques. PRIME+PROBE attacks have been performed in SEV-SNP \cite{wilke2023sev} which would greatly improve the resolution of the address space down to the cache line. With high enough resolution, handler functions would be separated across different cache line. This would enable distinguishing individual handler functions and inferring control flow at a much finer resolution. Combined with frequency analysis, akin to classical attacks on the Vigenère cipher, an attacker could recover the opcodes with great accuracy.\\
Prior work has shown that even high-level semantic content, such as images, can be reconstructed from page faults alone \cite{xu2015controlled}. Our attack improves the extraction of the confidential code deployed. This risk will only grow as new domains increasingly adopt Wasm. Recent work highlights the trend of deploying LLM inference workloads inside Wasm runtimes to enable secure, sandboxed, low-latency execution on edge devices \cite{zhang2025research}. Model inversion is a recent new threat in AI development and recent works has shown that model inversion attacks are more effective if the model structure is known prior to the attack \cite{xiang2020side}. The lack of code confidentiality would make it possible to infer this geometry which would make a subsequent attack stronger. Furthermore, the blockchain ecosystem such as Ethereum 2.0 continues shifting toward Wasm execution environments for smart contracts, where code confidentiality may directly affect financial integrity. Together, these trends motivate stronger protections for interpreter-based TEEs before such deployment becomes widespread.

\subsection{Mitigations}
\label{sec:discussion_mitigations}
Mitigating this class of attacks requires breaking structural regularities in interpreter behavior. One potential approach is to introduce multiple semantically equivalent handlers per opcode and randomize their selection, increasing the entropy of the observation surface. However, this increases maintenance complexity and may introduce new side-channel differences if not carefully balanced. More robust architectural protections exist, such as AEX-Notify \cite{constable2023aex} and TLBlur \cite{vanoverloop2025tlblur}, which hide page accesses and degrade instruction-level stepping. Unfortunately, these mechanisms currently target Intel SGX and rely on ISA extensions unavailable on AMD processors. \\
Obfuscation has traditionally been an approach to protect commercial intellectual property \cite{schrittwieser2016protecting}, and would be a potential means of protection. Adding randomness to the compile time of the WAMR interpreter would render our matching methodology futile. This could be done by inserting random \textbf{nop} operations, or shuffling around the code layout of the handlers. \\
Another direction is to abandon interpreter-style runtimes in favor of compilation to native code, thereby eliminating highly distinguishable handler dispatch patterns. While this effectively removes opcode-level leakage, it sacrifices several benefits of the Wasm ecosystem, including portability, fast startup, and robust sandboxing. Balancing security and practicality therefore remains an open design challenge for future confidential-computing platforms.

\section{Conclusion}
\label{sec:conclusion}
We have in this study shown the vulnerability of running an open source software such as an interpreter in regards to the exposed address space to a malicious hypervisor. Our pre-processing techniques, possible due to the exposed knowledge of the open source, result in significant features of the underlying runtime being exposed. This made it possible for us to extract a significant amount of executed code in a confidential setting, achieving recall values of 74.3\% across 128 million exported bytecode instructions. While similar work has been done on intel SGX, none has been conducted in a VM based TEE such as AMD SEV, and none to the same level of reconstruction as ours. 

\section*{Acknowledgment}
We would like to thank Luca Wilke for his quick responses and help regarding usage of the SEV-Step framework. He has on multiple occasions raised helpful insights on difficulties encountered when working on this study. Without his help, the work would have been much delayed.\\

This work was partially supported by the Wallenberg AI, Autonomous Systems and Software Program (WASP) funded by the Knut and Alice Wallenberg Foundation.\\

This article has received funding from the Smart Networks and Services Joint Undertaking (SNS JU) under the European Union’s Horizon Europe research and innovation program under Grant Agreement No 101139067. Views and opinions expressed are, however, those of the author(s) only and do not necessarily reflect those of the European Union. Neither the European Union nor the granting authority can be held responsible for them.
\printbibliography
\end{document}